\documentclass{PoS}

\usepackage[numbers]{natbib}
\usepackage{rotating}
\usepackage{graphicx}
\usepackage[english]{babel}
\usepackage[T1]{fontenc}
\usepackage[utf8]{inputenc}

\title{H.E.S.S. discovery of very-high-energy emission from the blazar PKS\ 0736+017: on the location of the $\gamma$-ray emitting region in FSRQs}

\ShortTitle{H.E.S.S. discovery of very-high-energy emission from the blazar PKS 0736+017}

\author{\speaker{Matteo Cerruti}\\
        Sorbonne Universités, UPMC Université Paris 06, Université
Paris Diderot, Sorbonne Paris Cité, CNRS, Laboratoire de
Physique Nucléaire et de Hautes Energies (LPNHE), 4 place
Jussieu, F-75252, Paris Cedex 5, France
\\
        E-mail: \email{mcerruti@lpnhe.in2p3.fr}}

\author{Jean Philippe Lenain\\
        Sorbonne Universités, UPMC Université Paris 06, Université
Paris Diderot, Sorbonne Paris Cité, CNRS, Laboratoire de
Physique Nucléaire et de Hautes Energies (LPNHE), 4 place
Jussieu, F-75252, Paris Cedex 5, France
}

\author{Heike Prokoph\\
GRAPPA, Anton Pannekoek Institute for Astronomy and
Institute of High-Energy Physics, University of Amsterdam,
Science Park 904, 1098 XH Amsterdam, The Netherlands
}

\author{for the H.E.S.S. Collaboration\\
}
\vspace{2cm}
\abstract{With the installation of a new 28-m diameter imaging atmospheric Cherenkov telescope
in the middle of the array, the H.E.S.S. instrument has entered since 2012 into its
Phase II. The fifth large-size telescope is particularly important to lower the threshold
energy of the array, and is thus a unique instrument to observe low-frequency-peaked
blazars, such as flat-spectrum radio-quasars (FSRQs), which remain rare in the very-high-
energy (VHE; E $>$ 100 GeV) $\gamma$-ray domain.  
In this contribution, we report on the discovery with the H.E.S.S. telescopes of VHE
$\gamma$-ray emission from the FSRQ PKS 0736+017 (z=0.189). H.E.S.S. observations were triggered
as a target-of-opportunity in February 2015 following the detection of $\gamma$-ray flaring activity
in the MeV-GeV energy-band with Fermi-LAT. Significant VHE emission is detected with
H.E.S.S. only during one of the nights of the observing campaign, showing at least night-by-night variability in the VHE regime. We discuss the location of the $\gamma$-ray emitting region within the relativistic jet, using opacity and variability constraints.    

}

\FullConference{35th International Cosmic Ray Conference – ICRC217-\\
		10-20 July, 2017\\
		Bexco, Busan, Korea}

\begin{document}

\section{Introduction}

The extragalactic sky at very-high-energies (VHE; E $>$ 100 GeV), as observed from the ground with imaging atmospheric Cherenkov telescopse (IACTs), is dominated by active galactic nuclei (AGN) of the blazar type. Blazars are considered, in the framework of the AGN unified model, as radio-loud AGN in which one of the two relativistic jets outflowing from the super-massive black-hole is closely aligned with the line of sight. The non-thermal emission from the jet, boosted by relativistic effects, overshines all other AGN emission component.\\ 
The blazar class is not homogeneous, and further divided into the two subclasses of Flat-Spectrum Radio-Quasars (FSRQs) and BL Lacertae objects (BL Lacs) according to the presence (in the former) or absence (in the latter) of broad emission lines in the optical/UV spectrum. Blazars are bright extragalactic emitters at all wavelengths, and observations have shown that blazars spectral energy distributions (SEDs) are comprised of two non-thermal components:  a low-energy one, peaking between millimeter and X-rays (in a $\nu F_\nu$ representation); and a high-energy one, peaking between MeV and TeV. The peak frequency of the low-energy component is used as an additional classification: while FSRQs are generally characterized by a peak frequency in infrared, BL Lacs show a variety of peak frequencies, and are further classied as low, intermediate, and high-frequency-peaked BL Lacs (LBLs, IBLs, HBLs). At VHE, the blazar population is dominated by HBLs, and low-frequency-peaked blazars as FSRQs and LBLs are much rarer.\\
Currently known VHE FSRQs are only a few: 3C 279, PKS 1222+216, PKS 1510-089, PKS 1441+25 and the gravitationally lensed blazar S3 0218+35 \citep{3C279magic, cerrutigamma, PKS1222magic, cerrutiintegral, 1510hess, 1510magic, 1441veritas, 1441magic, 0218magic}. All of them have been detected at VHE with Cherenkov telescopes only during flaring activity. In this contribution, we present the discovery with the H.E.S.S. telescopes of VHE emission from PKS 0736+017, a quasar at z=0.189, which thus becomes the sixth member of the small family of VHE FSRQs.\\
PKS 0736+017 is a very well studied quasar, first discovered as a powerful radio source with the Parkes telescopes \citep{Day66}. The source exhibits observational properties typical of blazars: its radio-morphology is characterized by a core and a single-sided jet; its optical/UV spectrum shows broad emission lines; its host galaxy is a giant elliptical. It is relatively nearby, being located at a redshift z = 0.189 \cite{redshift}, much closer than all other VHE FSRQs. The source experienced multiple flaring episodes in the past: in 2002, optical observations showed intra-night variability up to 0.6 mag/hour \cite{opticalflare}, which classifies the source as optically violently variable quasar. In $\gamma$-rays, the source is detected with Fermi-LAT, and included in the latest Fermi catalog \cite{3FGL} with a log-parabolic spectrum in the MeV-GeV band. Since the beginning of the Fermi-LAT mission, PKS 0736+017 remained quiet until November 2014, when it experienced a first bright flare \cite{atel}. After that, the source remained active, and experienced a second bright flare in February 2015, which triggered the H.E.S.S. observations reported in this contribution, and that resulted in the discovery of VHE emission.\\ 

H.E.S.S. is an array of five IACTs located on the Khomas Highland plateau, in Namibia (1800 m above sea level). Since 2012, the original configuration of four 12-m diameter telescopes has been complemented by a fifth 28-m telescope in the center of the array, marking the Phase II of the experiment. The fifth telescope provides a lower energy threshold, which is particularly suited for the study of low-frequency-peaked sources as FSRQs.\\
The telescopes observe the faint Cherenkov light emitted by electromagnetic cascades produced in the interaction of VHE photons with the Earth's atmosphere. Hadronic showers (produced by the interaction of cosmic rays with the Earth's atmosphere) are discriminated at the data-analysis stage by looking at the characteristics of the Cherenkov showers. The results presented in this contribution have been produced using the ImPACT analysis chain \cite{impact} and crosschecked using the model analysis chain \cite{mpp}.\\     
Data from H.E.S.S. Phase II can be analyzed in two ways: in the monoscopic, or the stereoscopic configuration. In the monoscopic approach, only events from the central 28-m telescope are used; in the stereoscopic mode, only events which triggered at least two telescopes are used. As presented in \cite{hessmono}, the monoscopic configurations provides a lower threshold, which is particularly useful for the study of soft sources.

\section{Data analyses}
\subsection{H.E.S.S. results}

\begin{figure}[t!]
\begin{center}
\includegraphics[width=200pt]{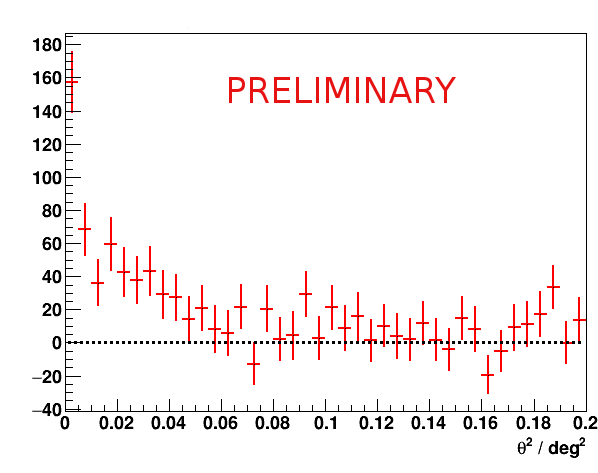}
\includegraphics[width=200pt]{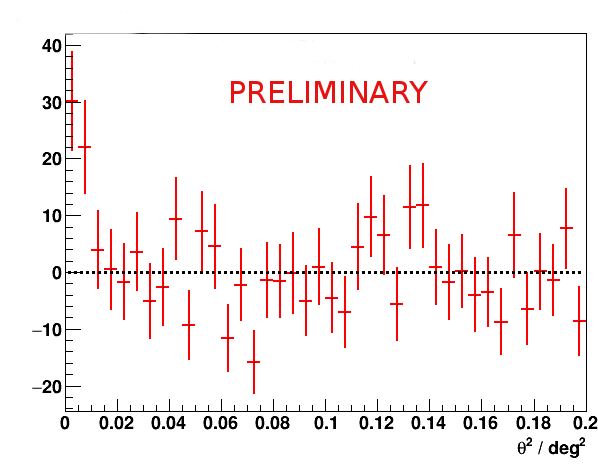}
\caption{Preliminary $\theta^2$ distributions of the excess at the position of PKS 0736+017, as measured with H.E.S.S. \textit{Left:} monoscopic reconstruction; \textit{Right:} stereoscopic reconstruction. \label{figimpact}}
\end{center}
\end{figure}

H.E.S.S. target of opportunity observations on PKS 0736+017 were triggered on Februrary 18, 2015 by the detection of $\gamma$-ray flaring activity using Fermi-LAT public data. H.E.S.S. observed the source on February 18, 19, and 21. Data were analyzed using both the monoscopic and the stereoscopic configurations: a point-like source of VHE photons at the position of PKS 0736+017 is detected during the night of February 19, 2015, only. During the other nights, only upper limits on the VHE emission could be put. In Fig. \ref{figimpact} we present the $\theta ^2$ excess  distributions for the night of Februrary 19, for both the monoscopic and stereoscopic reconstruction. The source is detected at 11 $\sigma$ and 5.5 $\sigma$ above the background, respectively.\\  
The spectral reconstruction is performed for the night of the detection, only. Both monoscopic and stereoscopic spectra are consistent with simple power-law emission, with no indication of cut-off. The photon index is estimated as $\Gamma = 3.1 \pm 0.3$ for the monoscopic analysis, and   $\Gamma = 4.2 \pm 0.8$ for the stereoscopic one. The errors represent the statistical uncertainty, only. It is important to underline that the two analyses have different thresholds, equal to 80 and 150 GeV, respectively. This can explain the different values of $\Gamma$ due to spectral curvature. The spectra are presented in Fig.\ref{figspectra}.\\
The night-by-night light-curve (for the monoscopic reconstruction, only, shown as integral flux above 100 GeV) is shown in Fig. \ref{figLC}. No evidence for intra-night variability is observed in the H.E.S.S. data during the night of February 19, 2015.

\subsection{Fermi results}

\begin{figure}[t!]
\begin{center}
\includegraphics[width=250pt]{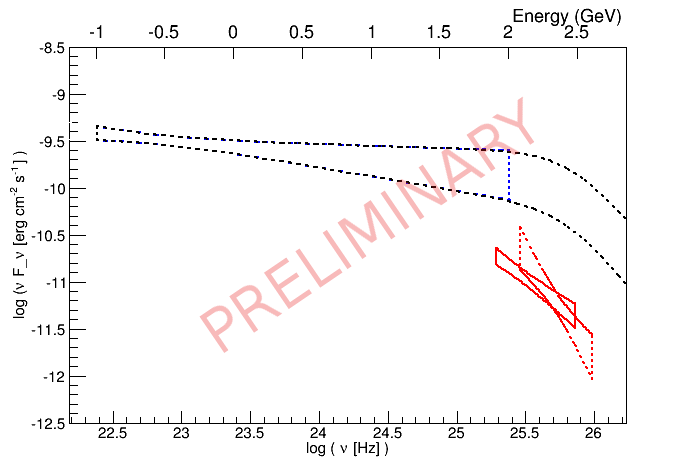}
\caption{Preliminary gamma-ray spectra of PKS 0736+017 measured with Fermi-LAT (in blue) and H.E.S.S. (in red; bold line for the monoscopic reconstruction, and dotted line for the stereoscopic reconstruction). The black dotted line represents the extrapolation of the Fermi-LAT spectrum towards higher energies assuming no cut-off and absorption on the extragalactic background light using the model by \cite{EBL}. \label{figspectra}}
\end{center}
\end{figure}

\begin{figure}[h!]
\begin{center}
\includegraphics[width=250pt]{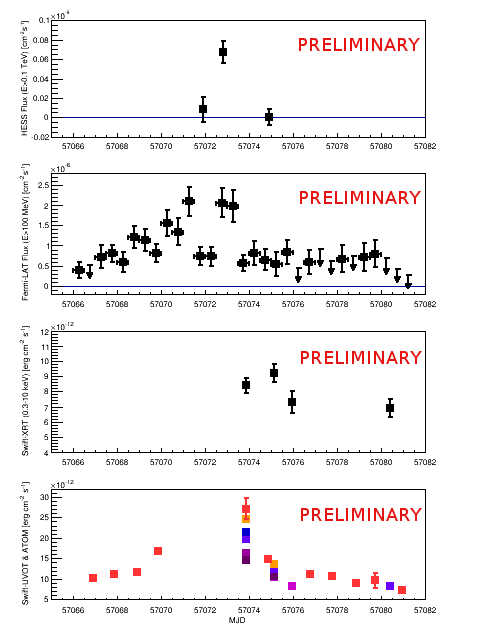}
\caption{Preliminary multi-wavelength light-curve of PKS 0736+017 during the February 2015 $\gamma$-ray flare. From top to bottom: H.E.S.S., Fermi-LAT, Swift-XRT, Swift-UVOT and ATOM.\label{figLC}}
\end{center}
\end{figure}

Fermi-LAT data on PKS 0736+017 have been analysed using Science Tools v10r0p5 and Pass8 instrument response functions. A first binned analysis is run on a broader time-range (February 1, 2015, to February 25, 2015) which encompasses the H.E.S.S. detection, to have a global view of the flare. The source is clearly detected within this data-set, with a significance level of 36$\sigma$. The spectrum is fitted with a power-law function with photon index $\Gamma = 2.2 \pm 0.05$, and integral flux (between 100 MeV and 500 GeV) = $(4.9 \pm 0.4)\times 10^{-10}$ erg cm$^{-2}$ s$^{-1}$. No evidence for spectral curvature has been found in the data. A light-curve with a time-binning of 12 hours has been computed assuming a power-law function, and letting the spectral-index free to vary. It is presented in Fig. 2. It shows that there is clear variability in the data, and that the $\gamma$-ray flare was indeed composed of two flares, the first one peaking on February 17, 2015, and the second one peaking on February 19, 2015, at the time of the H.E.S.S. observation window which yielded a detection.\\
In order to extract Fermi-LAT information contemporaneous with the H.E.S.S. observation on February 19, 2015, we performed a second Fermi-LAT analysis using only 24 hours of data from February 19, noon, to February 20, noon. PKS 0736+017 is again clearly detected, at a significance level of 18$\sigma$. The spectrum is compatible with the average value, with $\Gamma = 2.15 \pm 0.10$, and it is plotted in Fig. \ref{figspectra} together with the ones from the H.E.S.S. data analysis. \\

\subsection{X-ray and optical}

Following the detection of the VHE emission from PKS 0736+017 with H.E.S.S., target of opportunity observations with the Swift telescope have been carried out, using both the XRT and the UVOT instruments. Swift took four exposures on the source, on February 20, 22, and 27, 2015.\\
Swift-XRT data have been analyzed using Heasoft version 6.18. For each exposure, the spectrum is well fitted by a power-law function absorbed by Galactic material. For the latter, we used a value of $N_H = 7.8 \times 10^{20}$ cm$^{-2}$ \cite{nh}. The best-fit photon index is comprised between 1.34 and 1.62. For each XRT observation, the 0.3-10 keV unabsorbed flux from the spectral fit has been computed, and it is plotted in the lightcurve in Fig. \ref{figLC}. No evidence of variability is seen in the XRT data.\\   
Simultaneously with XRT, the Swift-UVOT instrument has observed PKS 0736+017, guaranteeing a coverage in the optical and UV band. During the first two Swift observations all six UVOT filters were used (V, B, U, W1, W2, M2), while during the third (respectively, fourth) observation only the W1 (respectively, U) filter was used. Fluxes have been extracted  using \textit{uvotmaghist}, and corrected for reddening using $E_{B-V} = 0.121$. The light-curve, in the bottom plot of Fig. \ref{figLC}, shows a decaying trend from February 20, to February 27, 2015.\\
The ATOM telescope is a fully automated optical telescope installed on the H.E.S.S. site. It regularly monitored PKS 0736+017, providing a long-term light-curve in the R band. The ATOM data are plotted in the bottom plot of Fig. \ref{figLC}. They show that an optical flare accompanied the $\gamma$-ray one, although no simultaneous coverage of the night of the H.E.S.S. detection is available.

\section{On the location of the $\gamma$-ray emitting region}

A major topic in the studies of $\gamma$-ray blazars is the location of the $\gamma$-ray emitting region. There is currently consensus in the community that high-energy photons are produced within the relativistic jet, but the details on the position (nearby the super-massive black hole, or downstream in the jet?), as well as on the radiative processes (leptonic or hadronic?) remain debated. An approach which proved to be succesfull in accessing the physics of $\gamma$-ray blazars is the modeling of the multi-wavelength spectral energy distribution. However, due to the intrinsic variability of blazars, this kind of study can be performed only in presence of strictly simultaneous observations across the electromagnetic spectrum. As shown in the previous sections, for PKS 0736+017 the only information we have during the H.E.S.S. detection on February 19, 2015, comes from Fermi-LAT, and no simultaneous data exist for longer wavelengths. In the following we thus make use only of $\gamma$-ray data to put constraints on the location of the emitting region during the H.E.S.S. detection. We make two explicit assumptions: the first one is that the radiative mechanism is leptonic, and in particular is inverse Compton scattering over an external photon field (the broad line region, or the dusty torus); the second one is that the emission from MeV to TeV is produced within a single emitting zone.

\subsection{Opacity constraint}

The first constraint that can be put using H.E.S.S. data only, is based on the opacity of the emitting region to $\gamma$-$\gamma$ pair production. As shown by several authors (see e.g. \cite{dp_abs, opacity}), if the $\gamma$-ray emission site is located at the jet basis, close to the black hole, the probability of an interaction between a VHE photon and a photon from the BLR is so high that a significant absorption effect should be seen in the VHE band. For PKS 0736+017, H.E.S.S. and Fermi-LAT contemporaneous observations show a break, which can be attributed to an intrinsic break of the $\gamma$-ray emission, or an absorption effect. By assuming that the break is due to absorption over BLR photons, the upper limit on the opacity is estimated to be $\simeq 2$, which sets thus a lower limit on the location of the emitting region r $> 8\times10^{16}$ cm, following \cite{opacity}.

\subsection{Other constraints}

\begin{figure}[t!]
\begin{center}
\includegraphics[width=250pt]{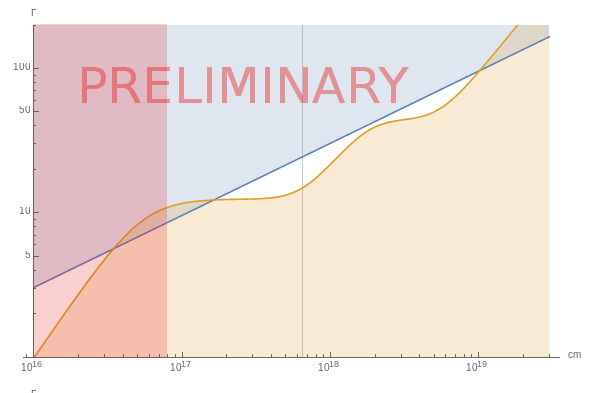}
\caption{Preliminary constraint on the location of the emitting region, as a function of the bulk Lorentz factor $\Gamma$. The red exclusion region represents the opacity constraint, 
the blue exclusion region represents the collimation constraint, while the orange excusion region represents the cooling constraint. The vertical black line represents the estimated location of the BLR.\label{figmodel}}
\end{center}
\end{figure}

Another constraint comes from the variability time-scale $\tau$, which has been measured to be $\simeq 12$ hours with Fermi-LAT. Under the assumption that the variability time-scale is related to the size of the $\gamma$-ray emitting region $R \simeq c \tau \frac{\delta}{1+z}$ (where $\delta$ is the Doppler factor of the emitting region), following \cite{nal}, we can set a limit based on the collimation of the emitting region $\Gamma * \theta$ < 1, where $\Gamma$ is the bulk Lorentz factor, and $\theta$ the opening angle. The limit represents an exclusion region in the $\Gamma$-r plane, and is shown in Fig. \ref{figmodel}.
An additional constraint comes from the cooling timescale $\tau_{c}$: under the assumption that the $\gamma$-ray emission is due to inverse Compton scattering of an external photon field, we can impose $\tau_{c} \leq \tau$, which, following \cite{nal} is translated into an exclusion region in the $\Gamma$-r plane.
The three above-mentioned constraints, using only the information from $\gamma$-ray observations, provide an estimation of the location of the emitting region during the VHE flare. The break in the $\gamma$-ray spectrum between Fermi-LAT and H.E.S.S allows for solutions located within the BLR. The minimum bulk Lorentz factor is estimated to be around 10.

\section{Conclusions}

We reported on the first detection of very-high-energy $\gamma$-ray photons from the FSRQ PKS 0736+017, which becomes the sixth FSRQ ever detected above 100 GeV so far. Multiwavelength observations with Fermi-LAT, Swift, and ATOM have been presented as well. Making use of $\gamma$-ray data only, we can pinpoint the location of the $\gamma$-ray emitting region, with a minimum bulk Lorentz factor of around 10.  Located at a redshift of 0.189, PKS 0736+017 is the nearest VHE FSRQ: multiwavelength observations during future bright $\gamma$-ray flares are encouraged.

\begin{small}
\section*{Acknowledgments} 
The support of the Namibian authorities and of the University of Namibia in facilitating the construction and operation of H.E.S.S. is gratefully acknowledged, as is the support by the German Ministry for Education and Research (BMBF), the Max Planck Society, the German Research Foundation (DFG), the Alexander von Humboldt Foundation, the Deutsche Forschungsgemeinschaft, the French Ministry for Research, the CNRS-IN2P3 and the Astroparticle Interdisciplinary Programme of the CNRS, the U.K. Science and Technology Facilities Council (STFC), the IPNP of the Charles University, the Czech Science Foundation, the Polish National Science Centre, the South African Department of Science and Technology and National Research Foundation, the University of Namibia, the National Commission on Research, Science and Technology of Namibia (NCRST), the Innsbruck University, the Austrian Science Fund (FWF), and the Austrian Federal Ministry for Science, Research and Economy, the University of Adelaide and the Australian Research Council, the Japan Society for the Promotion of Science and by the University of Amsterdam.
We appreciate the excellent work of the technical support staff in Berlin, Durham, Hamburg, Heidelberg, Palaiseau, Paris, Saclay, and in Namibia in the construction and operation of the equipment. This work benefited from services provided by the H.E.S.S. Virtual Organisation, supported by the national resource providers of the EGI Federation.\\
\end{small}

\bibliographystyle{aa}
\bibliography{PKS0736_biblio}

\end{document}